\documentclass{elsart}
\usepackage{natbib}
\usepackage{epsfig}
\begin{document}
\flushright{UCSD/PTH 00-15 \\ QMWC-HEP-2000-02}

\begin{frontmatter}
\title{The Glueball Spectrum}

\author[QMC]{D.V.~Bugg},
\author[SAND]{M.~Peardon},
\author[QMC]{B.S.~Zou\thanksref{Beijing}} \\
\address[QMC]{Queen Mary and Westfield College, London E1\,4NS, UK}
\address[SAND]{Dept. of Physics, UCSD, La Jolla, California 92093-0319, USA}
\thanks[Beijing]{Now at the Institute for High Energy Physics, Beijing
100039, China}

\begin{abstract}
Mass ratios of glueballs predicted by the
latest Lattice QCD calculations in the quenched approximation agree well with
four prime experimental candidates.
\end{abstract}
\end{frontmatter}

\newpage

There is an uncertainty of $\pm 10\%$ in the overall mass scale of
present Lattice QCD predictions for glueballs, arising from normalisation
to the mass of the $\rho$ meson and other benchmarks.
All simulations agree that the lowest $0^+$ glueball is to be expected in
the mass range 1500-1750 MeV \cite{1,2,3}.
This has led to suggestions by many authors that the $0^+$ glueball is related
to $f_0(1500)$ \cite{4,5}, or the nearby broad state $f_0(1530)$ \cite{6} or
$f_J(1710)$ \cite{3}.
The lowest $2^+$ and $0^-$ glueballs are predicted to lie close together
in the mass range 2000-2400 MeV.
A new feature of the latest calculations \cite{1} is that a second $0^+$
glueball is also predicted in the latter mass range. 

Mass {\it ratios} in the pure $SU(3)$ Yang-Mills theory are predicted
with greater accuracy. Predictions from Ref. \cite{1} are shown in the
second column of Table 1.  They agree remarkably well the the ratios
between $f_0(1500)$, $f_0(2105)$, $\eta (2190)$ and $f_2(1980)$ which
have exotic features, making them natural candidates for glueballs
\cite{7}. The lattice calculations of \cite{1,2,3} neglect the dynamics of the 
quark degrees of freedom and the lightest glueballs are stable,
physical states of the resulting purely gluonic theory. In QCD, the 
glueball states can both decay to light mesons and mix strongly with 
nearby $q\bar{q}$ resonances. The limitations of current simulations 
mean that little is known about the glueballs in lattice QCD including light 
quarks; the latest studies \cite{8,9} suggest a lower value
for the scalar mass. The mixing and decay of glueballs has been
addressed in the quenched approximation by the IBM group \cite{10,11}; 
they conclude that the scalar glueball is
rather narrow, with a total decay width to pseudoscalar pairs of
$\approx 100$ MeV and mixes strongly with the $s\bar{s}$ scalar
meson. We first review the candidates presented in Table 1, beginning
with $0^+$.

\begin{table}[htb]
\begin {center}
\begin{tabular}{cccc}
\hline
Ratio & Prediction & Experiment\\\hline
$M(2^{++})/M(0^{++})$ & 1.39(4) & 1.32(3) \\
$M(0^{-+})/M(0^{++})$ & 1.50(4) & 1.46(3)\\
$M(0^{*++})/M(0^{++})$ & 1.54(11) & 1.40(2)\\
$M(0^{-+})/M(2^{++})$ & 1.081(12) & 1.043(36)\\\hline
\end{tabular}
\caption{Glueball mass ratios predicted by Ref. \protect{\cite{1}},
compared with experimental candidates described in the text;
errors are in parentheses.}
\end {center}
\end{table}

A recent analysis of extensive data on $\bar pp \to \pi \pi$, $\eta \eta$
and $\eta \eta '$ has located many resonances in the mass range
1900--2400 MeV \cite{12}.
A straight-line trajectory of $0^+$ states against the square of the mass $M$
may be constructed, as shown in Fig. 1(a), with the same slope as is
observed for $2^+$ states, Fig. 1(b) and $4^+$.
(It is presently uncertain whether $f_0(980)$ is predominantly $\bar qq$
or a `molecule', but it does lie on the straight-line trajectory.)
The $f_0(1370)$, $f_0(1770)$, $f_0(2020)$ and $f_0(2320)$ all appear to decay
dominantly into non-strange states.
One expects $s\bar s$ states $\sim 250$ MeV above non-strange, and
$f_J(1710)$, which decays strongly to $K\bar K$, is a candidate for one
of them;
there is as yet no evidence for heavier $0^+~s\bar s$ resonances.
This leaves $f_0(1500)$ and $f_0(2105)$ as extra states.
Both have strong $\sigma \sigma$ decays and do not fit naturally as
$s\bar s$; $\sigma$ denotes the $\pi \pi$ S-wave amplitude.

The Particle Data Group (PDG) \cite{13} lists as established $I = 0$ resonances
only $f_0(980)$, $f_0(1370)$ and $f_0(1500)$, therefore
the evidence for remaining states will be reviewed here.
There is extensive evidence for $f_0(1770)$, listed by the PDG in 1996 as
$X(1740)$.
It appeared first in GAMS $\eta \eta$ data \cite{14} with a mass $M = 1744 \pm 15$
MeV; the observed decay angular distribution was flat, but $J^P = 2^+$
could not be excluded.
Next, the E765 collaboration \cite{15} observed striking peaks in $\eta \eta$
in $\bar pp \to (\eta \eta )\pi ^0$ at 1500, $1748 \pm 10$ and 2104 MeV,
but without a determination of $J^P$.
A similar series of peaks is visible in data from Mark III \cite{16} and
DM2 \cite{17}.
A re-analysis of these data finds all three peaks to have $J^P = 0^+$,
with decays mostly to $\sigma \sigma$ \cite{18}.
Data from the Crystal Barrel experiment on $\bar pp \to (\eta \eta )\pi ^0$
provide a good determination of the mass and width: $M = 1770 \pm 12$
MeV, $\Gamma = 220 \pm 40$ MeV \cite{19}.
These values leave little doubt that $f_0(1770)$ is distinct from the
$\Theta$, $f_J(1710)$.
Decay modes are also distinct: $\eta \eta$ and $\sigma \sigma$ for
$f_0(1770)$, but $K\bar K$ for $f_0(1710)$.

The $f_0(2105)$ was first identified in $J/\Psi \to \gamma (4\pi )$ \cite{18}.
It has recently been confirmed as a strong signal in $\bar pp \to \eta \eta$
\cite{12}.
There, it has been found that $\Gamma (\pi ^0\pi ^0)/\Gamma (\eta \eta ) =
0.71 \pm 0.17$, which is much too low for a normal non-strange $q\bar q$ state;
for the latter,
the predicted ratio is $(1/0.8)^4 = 2.45$, since the non-strange component of
the $\eta$ wave function has coefficient 0.8.
If the $f_0(2105)$ is treated as a mixed state
$\cos \theta (u\bar u + d\bar d)/\sqrt {2}
+ \sin \theta s\bar s$, this requires a mixing angle
$\theta = (65 \pm 6)^{\circ }$.
Its strong production from $\bar pp$ but dominant $\eta \eta$ decay suggests
exotic character.
It is also produced strongly in $\bar pp \to (\eta \eta )\pi ^0$ \cite{20}.
The PDG lists it under $f_2(2150)$.
Two further $0^+$ states have been reported: $f_0(2020)$ \cite{21,12} and
$f_0(2320)$ \cite{12}.

The $f_0(1500)$ is too close in mass to $f_0(1370)$ for both to be
explicable as $q\bar q$ states.
The existence of $f_0(1370)$ is therefore crucial.
It has been questioned by Minkowski and Ochs \cite{22}.
It is elusive because it decays weakly to $2\pi$, $K\bar K$
and $\eta \eta$, but dominantly into $4\pi$ .
A recent comparison of $\bar pp \to \eta (\pi ^0 \pi ^0)$ and
$\bar pp \to \eta (\pi ^+\pi ^- \pi ^+\pi ^-)$ \cite{23} establishes that
$\Gamma (4\pi )/\Gamma (2\pi ) \ge 5$, as does the analysis of $\bar pN \to
5\pi$ by Thoma \cite{24}.
It has been reported in several sets of data on $\bar pp \to 5\pi$ \cite{13}.
In the $2\pi$ channel, it has been observed in two independent analyses
of Crystal Barrel data on $\bar pp \to 3\pi ^0$ and $\eta \eta \pi ^0$
\cite{4,25} and also in $\bar pp \to (K\bar K)\pi ^0$ \cite{26}.
Its $\pi \pi$ width is sufficiently small that it cannot be identified
definitively in CERN-Munich data on
$\pi ^+\pi ^- \to \pi ^+\pi ^-$ \cite{27,28};
there, its peak cross section $\propto \Gamma ^2_{2\pi }/\Gamma ^2 _{total}$.
Nonetheless, it seems to be visible at large $|t|$ in GAMS data on
$\pi ^+\pi ^- \to \pi ^0\pi ^0$ \cite{29}.
It has been reported recently in central production of $2\pi$ \cite{30} and
$4\pi $ \cite{31}.

We turn now to $J^{PC} = 0^{-+}$.
A very broad signal with these quantum numbers is observed in
$J/\Psi \to \gamma (\rho \rho )$ \cite{16,17}.
Data on $J/\Psi $ radiative decays to $\rho \rho$, $\omega \omega$,
$K^*\bar K^*$ and $\phi \phi$ may be fitted with a single broad
resonance of mass $2190 \pm 40$ MeV having coupling constants which are
flavour-blind within experimental errors of $\sim \pm 30\%$,
i.e. in the ratio $3:1:4:1$ \cite{32}.
This is the classic glueball signature.

Lastly, we review $2^+$ states.
In Ref. \cite{12}, four $f_2$ were identified at 1920, 2020, 2210 and 2300 MeV.
Fig. 1(b) shows parallel
straight-line trajectories attributed to $\bar qq$ $^3P_2$
states $f_2(1270)$, $f_2(1565)$, $f_2(1920)$ and $f_2(2210)$ and
$^3F_2$ states $f_2(2020)$ and $f_2(2300)$.
The difference in mass
between $^3F_2$ and $^3P_2$ may be interpreted as arising from
the centrifugal barrier between $q$ and $\bar q$, which pushes $^3F_2$ up in
mass.
The $^3F_2$ states are almost degenerate with $^3F_4$ and $^3F_3$ \cite{33};
this is as expected, since the $L = 3$ centrifugal barrier shields the
short-range $L.S$ and tensor interactions.

In addition to these $f_2$'s, a broad $2^+$ state has been observed
with a mass of 1930--2000 MeV.
It appeared first in the data on central production of $4\pi$ \cite{34}.
It is also observed in $\bar pp \to (\eta \eta )\pi ^0$ as a broad
$\eta \eta$ signal with $M = 1980 \pm 50$ MeV \cite{20}.
There, one sees by eye a non-isotropic component in the decay angular
distribution across the whole mass range 1650--2200 MeV; such a broad state
cannot be explained by $f_2(1920)$.
A broad $2^+$ signal of similar mass and width is also reported in
$J/\Psi \to \gamma (K^*\bar K^*)$
\cite{35} and in $J/\Psi \to \gamma (4\pi )$ \cite{36}.
The latest data on Central Production of $4\pi$ final states shows a clear
broad peak with very little background \cite{31}; it has a mass of $1980 \pm 22$
MeV, with $\Gamma = 520 \pm 50$ MeV.
Recent WA102 Central Production data for $\omega \omega$ \cite{37} show it to be
distinct from $f_2(1920)$ of Fig. 1(b);
the latter has a conventional width of $\sim 200$ MeV.
Close, Kirk and Schuler \cite{38} show that $\phi$ distributions in
Central Production data are similar for $f_0(1500)$ and $f_2(1980)$ but quite
different to those for $f_0(1370)$ and $f_2(1270)$.
We interpret the $f_2(1980)$ as the $2^+$ glueball, probably
mixed with nearby $q\bar q$ states.

We now consider quantitative evidence concerning branching fractions for
production of glueballs in $J/\Psi \to \gamma (gg)$.
Close, Farrar and Li \cite{39} make predictions for branching fractions for
individual spin states depending on their masses and widths.
Using experimental values for these,
Table 2 makes a comparison of their predictions with experiment.
Their prediction is that $q\bar q$ states will be produced less strongly
than glueballs by a factor 5--10.

The branching fraction of $f_0(1500)$ observed in the $4\pi$ final
state \cite{18} is $(5.7 \pm 0.8) \times 10^{-4}$.
Our present best estimate of decay widths of $f_0(1500)$, slightly updated from
those of Ref. \cite{28}, are $\Gamma _{2\pi }: \Gamma _{K\bar K}: \Gamma _{\eta \eta
}:\Gamma _{4\pi } = 48:6:5:72$ MeV.
With these values, the total branching fraction of $f_0(1500)$ becomes
$(1.03 \pm 0.14) \times 10^{-4}$.
The result is close to prediction for a glueball and far above that
predicted for $q\bar q$, as pointed out in Ref. \cite{39}.
The production of $\eta (2190)$ in $\rho \rho$, $\omega \omega$,
$\phi \phi$, $K^*\bar K^*$, $\eta \pi \pi$ and $K\bar K\pi$ \cite{18,32} is
slightly above the prediction.
The $f_0(2105)$ has so far been observed in $J/\Psi$ radiative
decays only to $\sigma \sigma$ \cite{18}.
For this mode alone, the observed production is only 20\% of the prediction;
decays to $\eta \eta$ and $\pi \pi$ will also contribute,
but their decay branching ratios relative to $4\pi$ are not presently known.

The prediction for the branching fraction of the $2^+$ glueball is large
if the width is taken to be the 500 MeV fitted to  $f_2(1980)$.
Observed decays to $\sigma \sigma$ and
$f_2(1270)\sigma$ account for $(10 \pm 0.7 \pm 3.6)\times 10^{-4}$ of
$J/\Psi $ radiative decays \cite{36} and $K^*\bar K^*$ decays a further
$(7 \pm 1 \pm 2) \times 10^{-4}$ \cite{35}.
If one assumes flavour-blindness for vector-vector final states,
the vector-vector contribution  increases to $(16 \pm 2 \pm 4.5)\times 10^{-4}$.
The total of $2.6 \times 10^{-3}$
is still a factor 9 less than predicted for a glueball.
This is presently a major flaw in identifying $f_2(1980)$ with the
$2^+$ glueball.

It is possible that there are many decay modes as yet unobserved for the
heavy $f_0(2105)$ and $f_2(1980)$.
The total prediction for glueball production in $J/\Psi$ radiative decays
is 4.2\%, compared with $\sim 6\%$ for all radiative decays.
As yet, less than half of the products of $J/\Psi$ radiative decays
have been assigned to specific $J^P$.
Data on radiative decays to $\eta \eta$ and $\eta \eta \pi \pi$ would
be particularly valuable.

\begin{table}[htp]
\begin{center}
\begin{tabular}{ccc}
\hline
State & Prediction & Observation \\\hline
$f_0(1500)$ & $1.2 \times 10^{-3}$ & $(1.03 \pm 0.14)\times 10^{-3}$\\
$\eta (2190)$ & $1.4 \times 10^{-2}$ & $(1.9 \pm 0.3)\times 10^{-2}$\\
$f_0(2105)$ & $3.4 \times 10^{-3}$ & $(6.8 \pm 1.8)\times 10^{-4}$\\
$f_2(1980)$ & $2.3 \times 10^{-2}$ & $(2.6 \pm 0.6)\times 10^{-3}$\\\hline
\end {tabular}
\caption {Branching fractions of glueball candidates in $J/\Psi$ radiative
decays, compared with predictions of Close, Farrar and Li \protect{\cite{39}}.}
\end{center}
\end{table}

We now discuss decays and mixing with $q\bar q$.
Glueballs undoubtedly mix with neighbouring $q\bar q$ states.
The decay branching ratios of $f_0(1500)$ and $f_0(2105)$ are certainly
not flavour-blind \cite{5}.
A feature of $f_0(1500)$, $f_0(2105)$ and $f_2(1980)$ is that they
appear strongly in decays to $\eta \eta$ and $\sigma \sigma$.
The $\eta \eta$ decay is natural for a glueball, as pointed out by
Gershtein et al. \cite{40}.
Strohmeier-Prescicek et al. \cite{41} emphasize that the strong $\sigma \sigma$
decay points towards a large glueball component in $f_0(1500)$; analysing
observed decays, they deduce a 0.75 coefficient for the glueball component
in the wave function.
A major challenge now is to understand this mixing quantitatively.
In Ref. \cite{6} and further references cited there, the mixing of the
$0^+$ glueball with $q\bar q$ states has been fitted to an extensive range
of data.
Their conclusion is that $\sim 33\%$ of the glueball goes into $f_0(1500)$
and $\sim 53\%$ into a broad $0^+$ background, fitted as $f_0(1530)$ with
$\Gamma =1120 \pm 280$ MeV.

It is to be expected that glueballs will mix preferentially
with the ninth (predominantly
singlet) members of $q\bar q$ nonets.
We interpret the $\eta (1440)$ as the first radial excitation of
$\eta '(958)$.
Mixing between the broad $\eta (2190)$ and $\eta (1440)$ explains
naturally the strong production of $\eta (1440)$ in $J/\Psi$ radiative
decays and in $\bar pp$ annihilation \cite{32}.
The level repulsion between $\eta (2190)$ and $\eta (1440)$ can explain
the low mass of $\eta (1440)$ compared with $\eta (1295)$, $\pi (1300)$ and
$K_0(1460)$.
There is presently controversy whether $f_J(1710)$ has $J = 0$ or 2,
although recent publications by Dunwoodie \cite{42} and WA102 \cite{43}
favour spin zero. Its strong decay to $K\bar K$ suggests it is the ninth
member of a nonet having $J = 0$ or 2.
Mixing with the $0^+$ or $2^+$ glueball can explain its strong production
in $J/\Psi$ radiative decays.

Etkin et al. have observed anomalously strong production of $\phi \phi$
states with $J^P = 2^+$ in the mass range 2000-2340 MeV \cite{44}.
We conjecture that these are predominantly states of $s\bar s$ states
expected in this mass range, but enhanced strongly in $\pi \pi \to \phi \phi$
by mixing with the $2^+$ glueball, which couples to both initial and final
states. An $s\bar s$ $^3P_2$ state is expected
$\sim 250$ MeV above $f_2(1920)$ and could be responsible for the
S-wave $\phi \phi$ peak at 2150 MeV in Ref. \cite{44} and the peak observed
at the same mass in $K\bar K$ \cite{43}. A second $s\bar s$ $^3F_2$ state is
expected as partner to the $f_2(2020)$ and could explain the D-wave
peaks observed in $\phi \phi$ at 2300--2340 MeV.

Many puzzles and questions remain.
Why are the $\eta (2190)$ and $f_2(1980)$ so broad?  A partial answer
is that the glueball may mix with and
spread over many neighbouring $q\bar q$ states.
Nonetheless, the small width of $f_0(1500)$ is not understood.
Further progress on the mixing process requires an understanding the 
mixing process is required.

On the experimental side, it is vital to confirm (or contradict)
$f_2(1980)$ as the $2^+$ glueball. The glueball will couple
as the SU(3) singlet $u\bar u + d\bar d + s \bar s$.
The small $s\bar s$ content is hard to distinguish in
environments rich in $u\bar u +d\bar d$,
e.g. $\bar pp$ annihilation or pion-induced reactions.
Hence it is probably best studied in
$J/\Psi $ radiative decays and central production.
It is important to
test whether its decays to $\rho \rho$, $\omega \omega$, $\phi \phi$
and $K^*\bar K^*$ are flavour-blind. Present $J/\Psi$ data are statistically
weak and contain too much background at high masses to
allow a study of $\rho \rho$,
$\omega \omega$ and $\phi \phi$ channels.
It is also important to test flavour-blindness for decays of
$0^+$ glueball candidates to $\pi \pi$, $\eta \eta$ and $K\bar K$.
Presently, there are almost no data on $J/\Psi \to \gamma (\eta \eta )$.

In summary, although many puzzles remain, $f_0(1500)$, $f_0(2105)$,
$\eta (2190)$ and $f_2(1980)$ display exotic characteristics and do
not appear to fit naturally as $q\bar q$ states.
If they are identified as glueballs, or states strongly mixed with nearby
glueballs, mass ratios agree well with the latest predictions of
Lattice Gauge calculations of the glueball spectrum.

We acknowledge financial support from the
British Particle Physics and Astronomy Research Council (PPARC)
and the US DOE under grant No. DE-FG03-97ER40546. BSZ acknowledges
financial support from the Royal Society to visit QMWC.

\newpage

\begin{figure}
\begin{center}
\epsfig{file=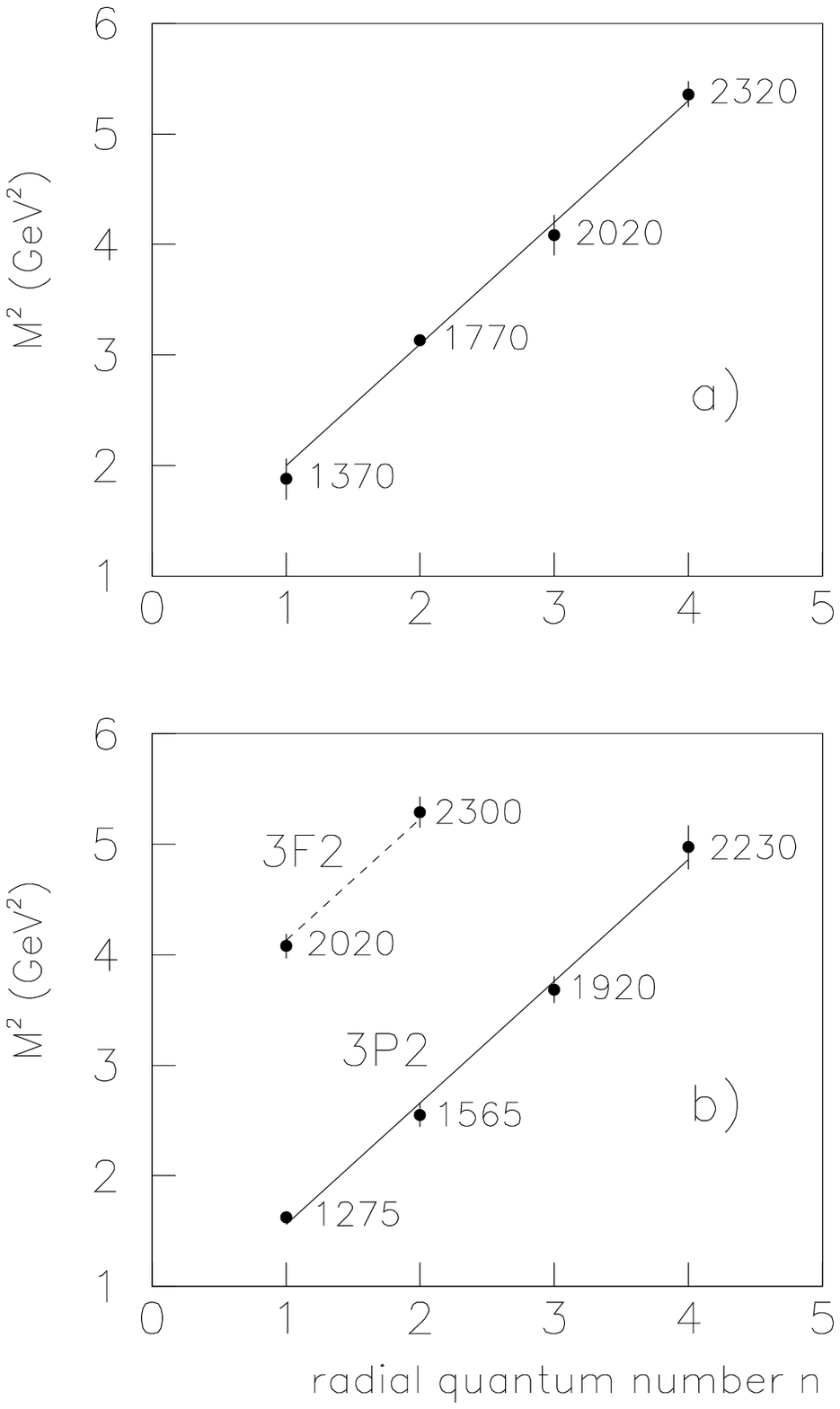,width=14.5cm}\
\caption{Trajectories v. $M^2$ for (a) $0^+$, (b) $2^+$ states;
numerical values give masses in MeV. }
\end{center}
\end{figure}
\end {document}